%% file: paper.tex
\documentclass[10pt,conference]{IEEEtran}
\IEEEoverridecommandlockouts
\usepackage{cite}
\usepackage{amsmath,amssymb,amsfonts}
\usepackage{algorithmic}
\usepackage{graphicx}
\usepackage{textcomp}
\usepackage{xcolor}
\usepackage{balance}
\usepackage[all]{nowidow}

\def\BibTeX{{\rm B\kern-.05em{\sc i\kern-.025em b}\kern-.08em
    T\kern-.1667em\lower.7ex\hbox{E}\kern-.125emX}}
\begin{document}

\title{Towards Predictive Replica Placement for Distributed Data Stores in Fog Environments\\
    \thanks{Funded by the Deutsche Forschungsgemeinschaft (DFG, German Research Foundation) -- 415899119.}
}

\author{\IEEEauthorblockN{Tobias Pfandzelter, David Bermbach}
    \IEEEauthorblockA{\textit{Technische Universit\"at Berlin \& Einstein Center Digital Future}\\
        \textit{Mobile Cloud Computing Research Group} \\
        \{tp,db\}@mcc.tu-berlin.de}
}

\maketitle

\begin{abstract}
    Mobile clients that consume and produce data are abundant in fog environments.
    Low latency access to this data can only be achieved by storing it in close physical proximity to the clients.
    Current data store systems fall short as they do not replicate data based on client movement.
    We propose an approach to predictive replica placement that autonomously and proactively replicates data close to likely client locations.
\end{abstract}


\input{sections/1_introduction}
\input{sections/2_architecture}
\input{sections/3_prediction}
\input{sections/4_conclusion}

\bibliographystyle{IEEEtran}
\bibliography{bibliography}

\balance

\end{document}

%% file: sections/1_introduction.tex
\section{Introduction}
\label{sec:introduction}

To reap the full potential of fog computing and enable emerging application domains such as the Internet of Things (IoT) or connected driving, platform architectures need to be redesigned for an increasing degree of geo-distribution.

In, e.g., a fog data store, full replication is infeasible as it requires constant communication between all nodes, especially when considering consistency guarantees.
Newer systems shard data and route client requests to the node that is concerned with a particular data item~\cite{MacCormick2009-rc}.
GFS~\cite{Ghemawat2003-gg} and Nebula~\cite{Ryden2014-ow} use centralized master servers that control replica placement.
This is practical in a tightly coupled cluster but routing requests to a central server negates any QoS improvements in a geo-distributed fog deployment.
Pastry~\cite{Rowstron2001-qi}, OceanStore~\cite{Kubiatowicz2000-he}, or Cassandra~\cite{Lakshman2010-zm} use hashing, which scales well and is easily implemented, but cannot take data movement based on proximity into account.

Instead, the close relationship between physical and virtual world in fog computing invites a new approach where data is sharded based on network topology and geographic distribution of application clients.
Most importantly, different clients rarely access the same data at different locations, e.g., an eHealth sensor is bound to a specific person and is independent from sensors on another person.

A fog data management system needs to ensure that the data the client needs is available at its closest node.
In iFogStor~\cite{Naas2017-ln}, a centralized cloud node calculates optimal data placement based on optimizing latency between clients and nodes, yet placement is static and as such not useful for mobile clients.
FBase~\cite{Hasenburg2020-yo,Hasenburg2019-oe} allows applications to control replica placement directly, which optimizes efficiency by moving data replicas with clients but places a burden on application developers.

We extend this approach with a component to predict application client movement that alleviates work needed on the application side.
The idea is that a client always accesses the same kind of data, so constantly moving this data to a node near the client leads to optimal resource allocation.
Crucially, this needs to happen without fully replicating all data, which is inefficient considering the limited network and storage resources available at the fog and edge.

%% file: sections/2_architecture.tex
\section{Architecture}
\label{sec:architecture}

Introducing a centralized component that manages replica placement limits scalability and introduces additional latency.
Furthermore, all access data would need to be relayed to this component, increasing stress on the network.

\begin{figure}
    \centering
    \includegraphics[width=0.9\columnwidth]{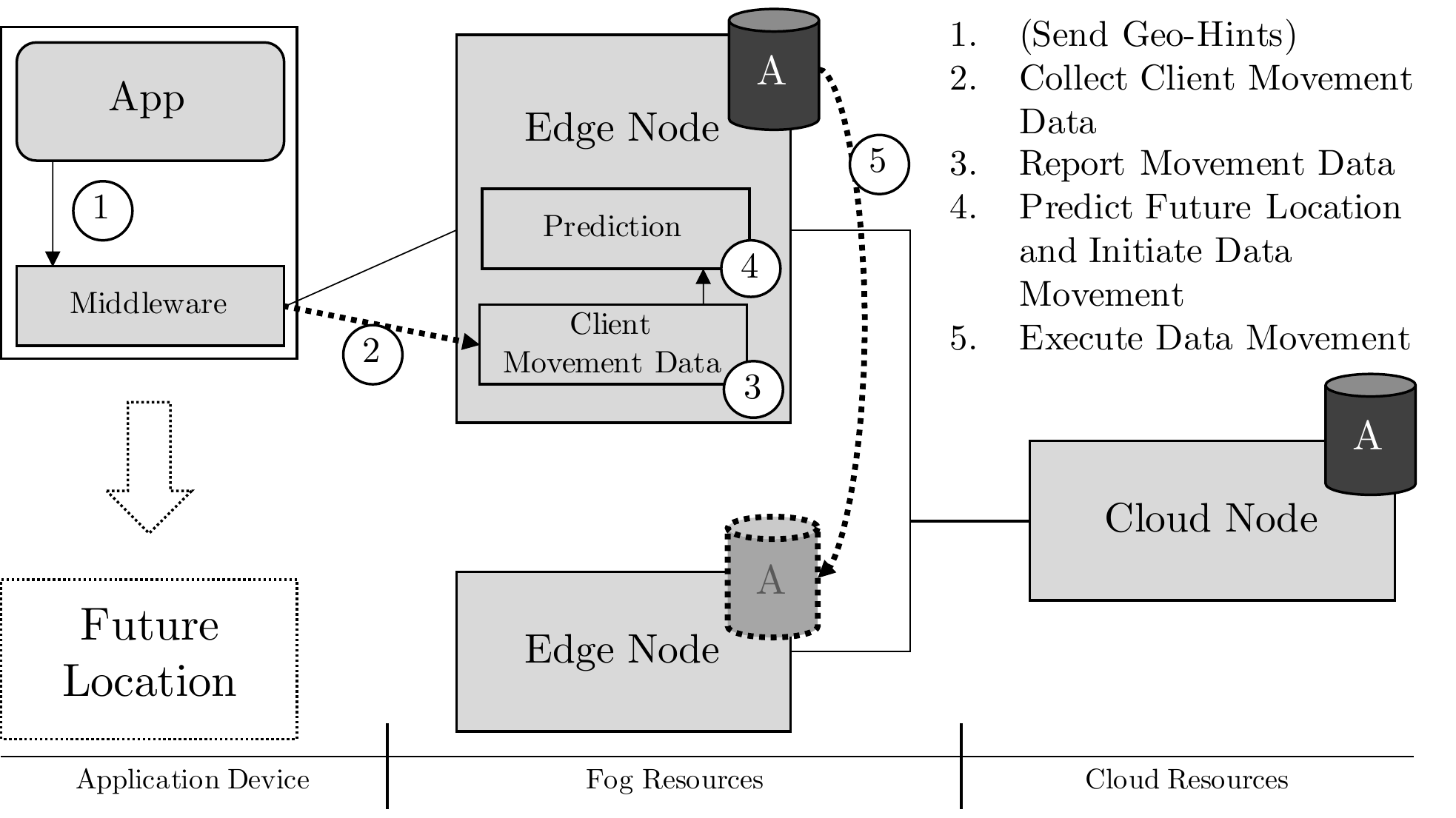}
    \caption{Hybrid Solution with Thin Client Middleware and Prediction Logic at Edge Nodes}
    \label{img:systemdesign}
\end{figure}

The better solution, then, is to decide replica placement at the edge in a decentralized manner.
We see two options for this: storage nodes keep track of data access and communicate among each other to coordinate data movement, or a client-side middleware initiates data movement.
The first approach enables thinner clients, can make use of resource pooling for different clients on the storage node, and can take topology information into account.
The second approach places a lesser burden on the edge node as prediction logic travels with the client, and sensitive information about data access and location can be kept on the client.

We propose a hybrid of both solutions.
As shown in Figure~\ref{img:systemdesign}, the client contains a small middleware that collects metadata about data access and, optionally, hints about future movement or data access from the application in the form of \textit{geo-hints}.
The middleware then relays the collected data to the edge node, which can also collect additional metadata.
This data is then used by a prediction component in the edge node that predicts future client location and initiates data movement.

Data about a particular client is kept mostly on the client so it does not have to be propagated throughout the storage system.
At the same time, executing prediction logic on the edge node enables more efficient clients.

%% file: sections/3_prediction.tex
\section{Predictive Replication Strategies}
\label{sec:solution}

When deciding whether to replicate a group of data items to a particular node, two costs have to be considered.
On the one hand, moving a replica incurs a cost for network traffic, increased complexity in managing consistency, and a cost for actually storing the data.
On the other hand, opportunity costs for not replicating have to be considered as well.
These could be increased access latency, decreased resiliency in case of node failure, or costs caused by missed QoS guarantees.

The optimal replication strategy for a data item requires knowledge of where and when application clients access this item.
Hence, predicting data placement means predicting data access.
This prediction can be based on the times a client accessed particular data in the past, on the physical or logical location of that access, on inferred metadata, or on client hints about future data access and location.

Client access statistics in the form of time series data can be collected.
From here, time series forecasting can help predict where the client might be located.
Movement of the client may also be considered.
When data shows that the client continuously moves between nodes in the same direction, future access is likely to happen at nodes further in this direction.
Metadata inferred from the data could, e.g., be the speed and direction of a connected car that it writes to the data store.
Finally, a client might have specific information on where it is moving and where future access will happen and can relay that information to the distributed storage system.

\begin{figure}
    \centering
    \includegraphics[width=0.75\columnwidth]{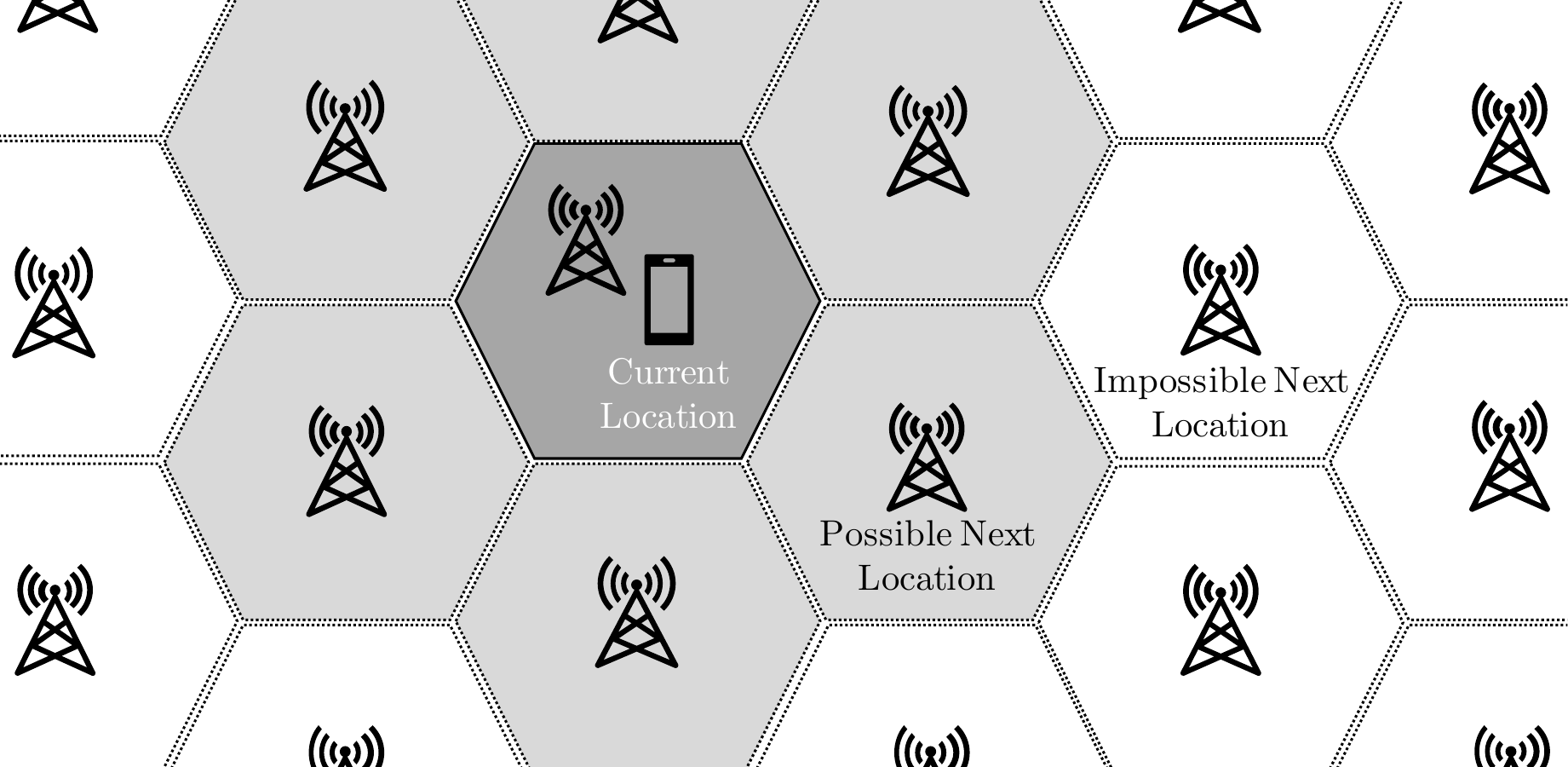}
    \caption{The Application Client Only Moves Between Adjacent Nodes and Cannot Skip a Node}
    \label{img:topology}
\end{figure}

Based on this data, time series forecasting, e.g., ARIMAX~\cite{Hyndman2018-wo} or machine learning might be used to predict future client location.
The prediction can be simplified when taking topology into account.
As shown in Figure~\ref{img:topology}, to reach the section of a node that is not adjacent to the client's current section, it must first pass through the section of an adjacent node.
Hence, when a node predicts the next location of the client, only adjacent nodes must be considered.
This greatly reduces possible locations and, consequently, nodes only need a local view of the system rather than knowledge of all other nodes.
Clients disconnecting and reconnecting at a different part of the world, e.g., a user with a mobile phone boarding a flight, are a particular challenge we plan to investigate.

Data movement within the system should also be transparent, as predicting location of application clients has privacy implications.
Metadata collected to realize this prediction could personally identify users and should as such be treated as personally identifiable data.
Predictions must also conform to data movement restrictions, e.g., a dataset might be required to not leave the European Union because of the GDPR.

%% file: sections/4_conclusion.tex
\section{Conclusion \& Future Work}
\label{sec:conclusion}

Current data store designs are ill-equipped to maximize efficiency in geo-distributed fog deployments.
Data should only be replicated close to the application clients that access it.
This replication should happen automatically to lower the barrier of fog computing adoption for application developers.
To that end, we present an approach towards predictive replica placement that predicts movement of application clients based on data such as past data access patterns and proactively replicates data to likely future locations.
This approach promises more efficient resource allocation in fog environments and optimal access latency to data for clients.

Our approach opens up interesting future research directions.
We plan to evaluate the effectiveness of different sources for data access and client location patterns, as well as different prediction methods.
Furthermore, we plan to compare the performance of our proposed system to other replica placement approaches, in particular with regards to access latency, overhead, and privacy in the context of data movement restrictions.

%% file: paper.bbl
\begin{thebibliography}{10}
\providecommand{\url}[1]{#1}
\csname url@samestyle\endcsname
\providecommand{\newblock}{\relax}
\providecommand{\bibinfo}[2]{#2}
\providecommand{\BIBentrySTDinterwordspacing}{\spaceskip=0pt\relax}
\providecommand{\BIBentryALTinterwordstretchfactor}{4}
\providecommand{\BIBentryALTinterwordspacing}{\spaceskip=\fontdimen2\font plus
\BIBentryALTinterwordstretchfactor\fontdimen3\font minus
  \fontdimen4\font\relax}
\providecommand{\BIBforeignlanguage}[2]{{%
\expandafter\ifx\csname l@#1\endcsname\relax
\typeout{** WARNING: IEEEtran.bst: No hyphenation pattern has been}%
\typeout{** loaded for the language `#1'. Using the pattern for}%
\typeout{** the default language instead.}%
\else
\language=\csname l@#1\endcsname
\fi
#2}}
\providecommand{\BIBdecl}{\relax}
\BIBdecl

\bibitem{MacCormick2009-rc}
J.~MacCormick, N.~Murphy, V.~Ramasubramanian, U.~Wieder, J.~Yang, and L.~Zhou,
  ``Kinesis: A new approach to replica placement in distrib. storage systems,''
  \emph{Trans. Storage}, vol.~4, no.~4, pp. 11:1--11:28, 2009.

\bibitem{Ghemawat2003-gg}
S.~Ghemawat, H.~Gobioff, and S.-T. Leung, ``The {Google} file system,'' in
  \emph{Proc. Nineteenth ACM Symp. Operating Syst. Principles}, Oct. 2003, p.
  29–43.

\bibitem{Ryden2014-ow}
M.~Ryden, K.~Oh, A.~Chandra, and J.~Weissman, ``Nebula: Distrib. edge cloud for
  data intensive computing,'' in \emph{2014 {IEEE} Int. Conf. Cloud
  Engineering}, Mar. 2014, pp. 57--66.

\bibitem{Rowstron2001-qi}
A.~Rowstron and P.~Druschel, ``Pastry: Scalable, decentralized object location,
  and routing for large-scale peer-to-peer systems,'' in \emph{Middleware
  2001}, Oct. 2001, pp. 329--350.

\bibitem{Kubiatowicz2000-he}
J.~Kubiatowicz, D.~Bindel, Y.~Chen, S.~Czerwinski, P.~Eaton, D.~Geels,
  R.~Gummadi, S.~Rhea, H.~Weatherspoon, W.~Weimer, C.~Wells, and B.~Zhao,
  ``{OceanStore}: An architecture for global-scale persistent storage,''
  \emph{SIGARCH Comput. Archit. News}, vol.~28, no.~5, pp. 190--201, 2000.

\bibitem{Lakshman2010-zm}
A.~Lakshman and P.~Malik, ``Cassandra: A decentralized structured storage
  system,'' \emph{Oper. Syst. Rev.}, vol.~44, no.~2, pp. 35--40, 2010.

\bibitem{Naas2017-ln}
M.~I. Naas, P.~R. Parvedy, J.~Boukhobza, and L.~Lemarchand, ``{iFogStor}: An
  iot data placement strategy for fog infrastructure,'' in \emph{2017 {IEEE}
  1st Internat. Conf. Fog and Edge Comput. ({ICFEC})}, May 2017, pp. 97--104.

\bibitem{Hasenburg2020-yo}
J.~Hasenburg, M.~Grambow, and D.~Bermbach, ``Towards a replication service for
  data-intensive fog applications,'' in \emph{Proc. 35th {ACM} Symp. Appl.
  Computing, Posters Track ({SAC} 2020)}, Apr. 2020.

\bibitem{Hasenburg2019-oe}
------, ``{{FBase}: A Replication Service for {Data-Intensive} Fog
  Applications},'' Tech. Rep., 2019.

\bibitem{Hyndman2018-wo}
R.~J. Hyndman and G.~Athanasopoulos, \emph{Forecasting: Princ. and
  Practice}.\hskip 1em plus 0.5em minus 0.4em\relax OTexts, 2018.

\end{thebibliography}
